\documentclass[runningheads]{llncs}
\input{psfig.sty}
\begin{document}

\pagestyle{headings}
%In order to omit page numbers and running heads
%please change this line to
%\pagestyle{empty}
%and change the first command line too, see above.

\mainmatter

\title{Short- and Long-Term Statistical Properties of
Heartbeat Time-Series
in Healthy and Pathological Subjects}

\titlerunning{Lecture Notes in Computer Science}

\author{Paolo Allegrini\inst{1}
\and Rita Balocchi\inst{2} \and
 Santi Chillemi\inst{3} \and
Paolo Grigolini\inst{3,4,5} \and Luigi Palatella\inst{4} \and
Giacomo Raffaelli\inst{6}}

\authorrunning{Paolo Allegrini et al.}

\institute{Istituto di Linguistica Computazionale 
del Consiglio Nazionale delle
Ricerche,
Area della Ricerca di Pisa-S. Cataldo,
Via Moruzzi 1,
56124, Pisa, Italy.
\email{(paolo.allegrini@ilc.cnr.it)},
\and
Istituto di Fisiologia Clinica 
del Consiglio Nazionale delle
Ricerche,
Area della Ricerca di Pisa-S. Cataldo,
Via Moruzzi 1,
56124, Pisa, Italy.
\and
Istituto di Biofisica
del Consiglio Nazionale delle
Ricerche,
Area della Ricerca di Pisa-S. Cataldo,
Via Moruzzi 1,
56124, Pisa, Italy
%\email{\{Hofmann, Beyer, Kramer, Erika.Siebert-Cole, LNCS\}@Springer.de}\\
%\texttt{http://www.springer.de/comp/lncs/index.html}
%\email{Buerk@Springer.de}}
\and
Dipartimento di Fisica dell'Universit\`{a} di Pisa and INFM, via
Buonarroti 2, 56127 Pisa, Italy.
\and
Center for Nonlinear Science, University of North Texas,
P.O. Box 311427, Denton, Texas, 76203-1427 USA.
\and
International School for Advanced Studies and INFM Unit,
via Beirut 2-4, 34014 Trieste, Italy.}

\maketitle

\begin{abstract}
We analize heartbeat time-series corresponding to several groups of
individuals (healthy, heart transplanted, with congestive heart
failure (CHF), after myocardial infarction (MI), hypertensive),
looking for short- and long-time statistical behaviors. In particular
we study the persistency patterns of interbeat times and
interbeat-time variations. Long-range correlations are revealed using
an information-based technique which makes a wise use of the available
statistics. The presence of strong long-range time correlations seems
to be a general feature for all subjects, with the exception of some
CHF individuals. We also show that short time-properties
detected in healthy subjects, and seen also in hypertensive and MI
patients, and completely absent in the trasplanted, are characterized
by a general behavior when we apply a proper coarse-graining procedure
for time series analysis.
%when time series undergo a coarse-graining
%procedure.

\end{abstract}

\section{Introduction}

In the last few years, time series analysis
techniques have been applied to heartbeat sequences
to detect long-range memory \cite{shlesinger,stanley,rita}.
With these methods
one can in principle distinguish between different
statistical properties that may correspond to  different
pathophysiological condition.
Another important goal of this analysis
is to unravel the hidden control mechanisms
responsible for the heartbeat dynamics.
Indeed heart rate variability is influenced by the
``competition'' between sympathetic and parasympathetic
nervous system activity as well as by non-autonomic factors.
Consequently, an analysis regarding different time scales can
in principle disentangle the different contributions
to the heart rate variability.
In this work we use a recently developed
technique for measuring long-range correlations called
{\em Diffusion Entropy} (DE) \cite{giacomo,granada,PRE}.
We also focus on short-range time behavior
and, using a coarse-graining procedure, we
identify an objective resolution time for the measure
of the interbeat distance. We use this results to
compute the relative importance of the short-time
dynamics with respect to the long-term one.

\section{Materials and methods}

\subsection{The diffusion entropy technique}

In order to process the data, the first step
consists of a coarse-graining procedure on the RR time series.
This strategy, shared also by other groups
(see, for instance, Ref. \cite{stanley}) makes a statistical
treatment of the data possible.

We define a coarse-graining parameter $s$ and we
obtain a new series, namely
\begin{equation}\label{CGprocedure}
T_i(s) \equiv [T_i/s],
\end{equation}
where $T_i$ is the time distance between the $i^{th}$ and
the $i-1^{th}$ beat, $T_i(s)$ is the coarse-grained time series
and $[\cdot]$ denotes the integer part.
Equation (\ref{CGprocedure}) means that we
preliminary divide the interbeat distance scale
in several ``boxes'' of size $s$ and we assign
to all the interbeat distances lying
in the same ``box'' the same value in the sequence $T_i(s)$.
Next, we convert this series into a dichotomous sequence,
setting $\xi_i=1$ when $T_i(s) \neq T_{i-1}(s)$, and $\xi_i=0$
if $T_i(s) = T_{i-1}(s)$.
Finally, we generate several trajectories
(labelled with the index $l$) for
the variable $x_l$ at ``time'' $t$, namely,
\begin{equation}
x_l(t)=\sum \limits_{i=l}^{l+t} \xi_i.
\end{equation}
Note that, for simplicity, we have omitted indicating
the dependence on $s$.
As shown in Ref. \cite{giacomo}, if the
sequence $\xi_i$ is ergodic, then
the probability distribution of the variable
$x$ as a function of $t$ is
expected to fit the scaling property
\begin{equation}
p(x,t) = \frac{1}{t^{\delta}}F\left(\frac{x}{t^{\delta}}\right),
\label{scaling}
\end{equation}
with the ``degree of anomaly'' being measured by the
distance of the scaling parameter $\delta$
from the standard value $0.5$ corresponding to Brownian motion.

It is straightforward to prove that
the Shannon entropy
\begin{equation}
S(t) = - \int_{-\infty}^{\infty} p(x,t) \ln (p(x,t)) {\rm d}x
\label{shannonentropy}
\end{equation}
of a process fitting the scaling
condition of Eq. (\ref{scaling})
 yields
\begin{equation}
S(t) = A + \delta \ln (t) ,
\label{entropyform}
\end{equation}
where $A$ is a constant, whose explicit form is not relevant for
the ensuing discussion.  This result is immediately obtained by
plugging Eq. (\ref{scaling}) into Eq. (\ref{shannonentropy}).
The above method of evaluating the scaling parameter $\delta$ 
is called Diffusion Entropy (DE), and, as it will become clear later, this technique
is more efficient
than the calculation of the second moment of the probability distribution.
Note that
when the distribution density under study departs from the ordinary
Gaussian case and the function $F(y)$ has slow tails with an inverse power
law nature \cite{giacomo} the second moment
is a divergent quantity. This diverging quantity is made finite by the
unavoidable statistical limitation. In this case, the second moment
analysis 
would be determined by the statistical inaccuracy, thereby
leading to misleading conclusions, while the method based on  
Eq. (\ref{entropyform}) yields correct results \cite{giacomo,granada}.
%would yield misleading results, determined by
%the statisticaly inaccuracy,
%while the method based on Eq. (\ref{entropyform}) yields
%correct results \cite{giacomo,granada}.

\subsection{Short-time analysis}

Let us define the sequence $\{\tau_j\}$ as the distance
between $1$'s in the series $\{ \xi_i \}$.
The authors of Ref. \cite{giacomo} studied the case where
the probability distribution for $\tau$ decays as an inverse-power
law with exponent $\mu>2$.
If $\tau_j$ were uncorrelated, then
a simple expression would exists, linking $\mu$ and $\delta$, namely
\begin{equation}
\label{nomemory}
\delta = \frac{1}{\mu -1} \mbox{ } (\delta = 0.5 \mbox{ if } \mu > 3).
\end{equation}
In our analysis we observe a strong discrepancy
between the values of $\delta \approx 0.8$ and of
$\mu \approx 5$, thus suggesting that
the sequence of $\tau_j$ has a long time memory \cite{PRE}.
This is why we perform the normalized auto-correlation
function of the sequence $\tau_j$ defined as
\begin{equation}
\Phi_{\tau}(j)=\frac{\left \langle (\tau_k-\langle \tau \rangle) (\tau_{k+j}-\langle \tau \rangle) \right \rangle}
{\langle \tau_k^2 \rangle-\langle \tau \rangle^2}
\end{equation}
where $\langle \cdot \rangle$ is the arithmetic average, e.g.
\begin{equation}
\langle \tau \rangle \equiv \frac{1}{N} \sum\limits_{k=0}^{N} \tau_k ,
\end{equation}
and $N$ is the number of terms in the sum.

At this point we realize that for
most data the correlation function $\Phi_{\tau}(j)$
actually decays as an inverse power law after an abrupt
fall occurring at the transition from $j=0$ to $j=1$. The value at
$j=1$, defined as $\epsilon^2$, exhibits a peculiar
behavior as a function of the coarse-graining parameter $s$.
We will show that this behavior is typical of all
the groups, with the exception of transplanted patients and
some CHF individuals.

\subsection{RR time series}

The RR series analyzed are
taken from the { \em NOnLinear TIme Series AnaLysIS}
(NOLTISALIS) archive.
This data set is the result of the collaboration of several
interdisciplinary Italian research centers.
The data correspond to 50 patients classified in 5 groups
of 10 subjects each:
normal, hypertensive, post-MI,
with congestive heart failure, and heart transplanted.

For each subject there are 24-hour Holter recordings,
whose sample frequencies range
from 128 to 1024 Hz, according to the type of pathology.

\section{Results}
\subsection{Long-term analysis}

\begin{figure}
%\label{fig1}
\centerline{\psfig{figure=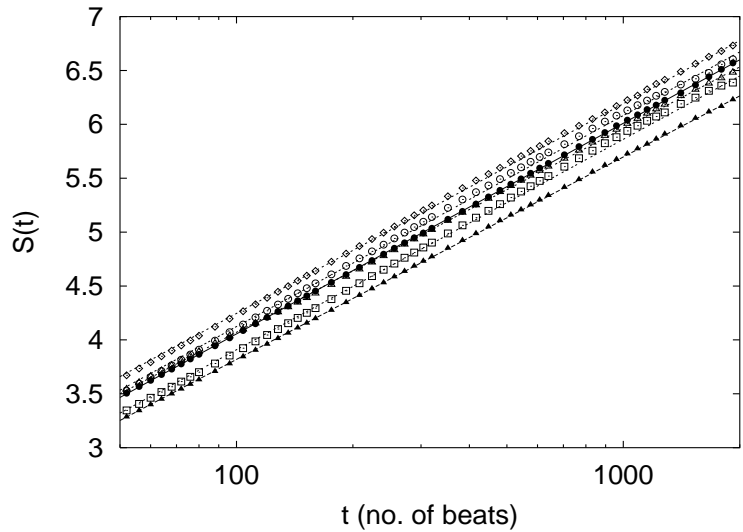,width=10.5cm}}
\caption{Diffusion Entropy S(t) for healthy patients. The straights
lines in log-linear scale represent best fits for Eq. (5)}
\label{fig1}
\end{figure}

\begin{figure}
%\label{fig2}
\centerline{\psfig{figure=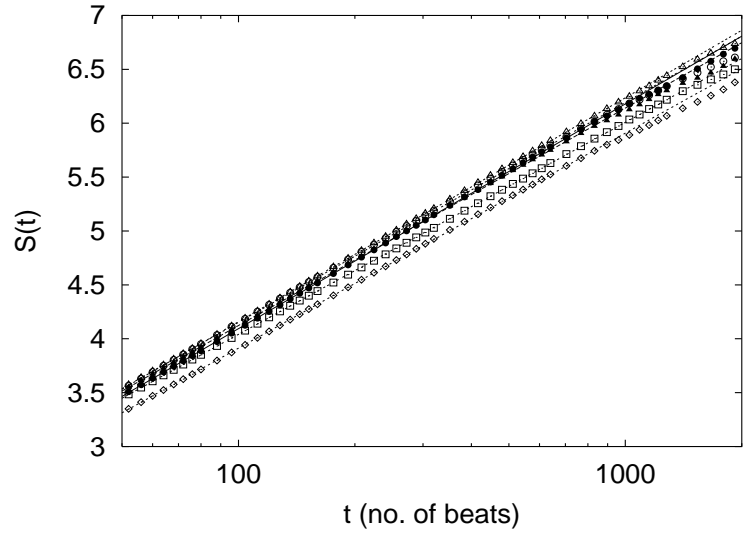,width=10.5cm}}
\caption{Diffusion Entropy S(t) for patients hypertensive. The straights
lines represent best fits for Eq. (5)}
\label{fig2}
\end{figure}

The analysis performed with the Diffusion Entropy technique
yields very accurate fits as shown in Figs. \ref{fig1} and \ref{fig2}. The average
results for each group of patients is shown in table \ref{tab1}.

\begin{table}
\begin{center}
\caption{Average value of $\delta$ for each
group of patients \label{tab1}}
\begin{tabular}{|c|c|}
\hline
{\bf Group} & $\langle \delta \rangle$ \\
\hline
healthy & $0.83 \pm 0.02$ \\
CHF & $0.80 \pm 0.05$ \\
post-MI & $0.85 \pm 0.04$ \\
transplanted & $0.86 \pm 0.08$ \\
hypertensive & $0.87 \pm 0.02$ \\
\hline
\end{tabular}
\end{center}
\end{table}

As shown in table \ref{tab1}, the values of $\delta$
indicate very strong correlations for all the different groups.
The CHF individuals, however, as previously shown in Ref. \cite{stanley,PRE},
are characterized by a somewhat smaller value.
%However, the chf individuals previously studied by
%us and other groups in \cite{stanley,PRE} belong to
%a different data set and present a consideable slower value for $\delta$.
One of the main results of this paper is that, remarkably, the results
from table \ref{tab1} are independent
of the coarse-graining parameter $s$.
As we shall show later, this property can be
accounted for by a simple intermittent model.

On the other hand, the coarse-graining parameter $s$ becomes very important
in evaluating
the correlation function $\Phi_{\tau}(j)$.
In fact, different values of $s$ yield different
series for $\tau$, while
a large/small coarse graining leads to
higher/lower value for $\langle \tau \rangle$.
We investigate the different correlation functions
stemming from different coarse-graining values $s$. As earlier stated, in most cases
the correlation function presents an abrupt drop going
from $j=0$ to $j=1$. After this fast drop, it decays to zero very slowly,
like an inverse-power law.
In fact, the data are not clean enough to establish
the inverse power law index with a good accuracy.
This means that
the DE is a much more efficient method to detect memory and much less ambiguous
than the correlation function, and we can
actually describe the meaning of $\delta$ making use of a general
class of dynamical models.
The correlation function $\Phi_{\tau}(j)$, on the other hand,
depends on the details of the model and does not
afford an easy way to derive it.

%\subsection{Short term analysis}
The numerical form of the correlation function $\Phi_{\tau}(j)$ is
        \begin{equation}
         \Phi_{\tau}(j) = (1-\epsilon^2) \delta_{j,0} +\epsilon^2 C(j).
         \label{cmm}
         \end{equation}
Here $\delta_{j,0}$ denotes the Kronecker $\delta$ step,
while the function $C(j)$
is a smooth function with the property $C(0)=1$.
We account for the long-term properties, 
namely second term in the r. h. s of (\ref{cmm}),
as follows.
We hypothesize that the heartbeat rate, on a rather long time scale,
undergoes trends of acceleration and deceleration of slopes $\alpha$,
whose duration length $\tau$ are asymptotically 
distributed with an inverse power law,
with index $\mu'$.
We suppose that the consecutive trends are not correlated and
consequenlty the scaling parameter $\delta$
is connected to $\mu'$ via \cite{giacomo}
\begin{equation}\label{deltamuprimo}
\delta=\frac{1}{\mu'-1}.
\end{equation}
Here we sketch a proof of Eq. (\ref{deltamuprimo}). First of all
we define the distribution probability for the slope $\alpha$, $f(\alpha)$.
We make the simplifying and plausible hypothesis that this distribution 
has no infinite moments.
This restriction does not apply to the $\tau$-distribution,
which has, on the contrary slow tails and diverging moments. 
Without loss of generality we assume the following form
\begin{equation}
\psi(\tau)=\frac{A^{\mu'-1}}{\left (A+\tau \right )^{\mu'}},
\end{equation}  
where $A=\langle \tau \rangle/(\mu'-2)$.
During one single trend of acceleration or deceleration
the position of the walker $x_l(t)$ is displaced by a quantity
\begin{equation}
z=\frac{|\alpha|}{s} \tau,
\end{equation}
which is the number of times the walker crosses the ``box boundaries''
defined by the coarse-graining procedure on $T_i$.
Consequently, we can write the probability of a displacement $z$
that reads
\begin{equation}
p(z)=\int\limits_{0}^{\infty} {\rm d}\tau \int\limits_{-\infty}^{\infty} {\rm d}\alpha 
\delta \left (  z-\frac{\alpha \tau}{s} \right ) \psi(\tau) f(\alpha).
\label{strangeconvolution}
\end{equation}
After many trends it is straightforward to see, e. g. along the 
lines of Ref. \cite{mario}, 
that the movement of our walker
is essentially equivalent 
to a process where a walker, at regular time distances
of duration $\langle \tau \rangle$, 
%to a L\'evy flight, i.e. a random
%walk where at each time step at a distance $\langle \tau \rangle$ 
%the walker jumps 
walks by a quantity
$\Delta x \sim z$, distributed as
\begin{equation}
\Pi(\Delta x)=\frac{s}{|\alpha| \langle \tau \rangle}\psi \left( \frac{s \Delta x}{|\alpha|} \right).
\label{psideltaics}
\end{equation}
Since the function $\psi$ has a long tail, this latter process, according
to the Generalized Central Limit Theorem \cite{gnedenko}, results in a L\'evy
Flight, which is a stable process with diverging central moments.  
The difficulty of dealing directly with the distribution (\ref{psideltaics})
is due to the fact that $\alpha$ is itself a stochastic variable.
However we can detect the asymptotic scaling property of the walker $x_l(t)$ 
by studying the first diverging moment of the distribution $p(z)$.
To find this value we evaluate
\begin{equation}
  \langle z^{\gamma} \rangle= \int\limits_0^{\infty}  p(z) z^{\gamma}{\rm d}z. 
\label{zetatogamma}
\end{equation}
The smallest value of $\gamma$ for which this integral diverges, called $\hat{\gamma}$, defines the scaling $\delta$ via $\delta=1/\hat{\gamma}$ \cite{gnedenko}.
Plugging Eq. (\ref{strangeconvolution}) into (\ref{zetatogamma}), we obtain
\begin{equation}
\langle z^{\gamma} \rangle= \int\limits_0^{\infty}  {\rm d} z
\int\limits_{0}^{\infty} {\rm d}\tau \int\limits_{-\infty}^{\infty} {\rm d}\alpha 
\:\: \delta \left (  z-\frac{\alpha \tau}{s} \right ) \psi(\tau) f(\alpha) z^{\gamma}.
\end{equation}
Integrating over $z$, this result can be written as
\begin{equation}
\langle z^{\gamma} \rangle=\left( \int\limits_0^{\infty} \psi(\tau) \tau^{\gamma}{\rm d}\tau  \right) \cdot F_s,
\end{equation}
where
\begin{equation}
F_s=\int\limits_{-\infty}^{\infty} {\rm d}\alpha 
\left ( \frac{|\alpha|}{s} \right )^{\gamma} f(\alpha).
\end{equation}
This means that only the first factor might make $\langle
\tau^{\gamma} \rangle$ divergent, thereby implying that
$\hat{\gamma}=\mu'-1$ and proving Eq. (\ref{deltamuprimo}).  This also
proves that $\delta$ {\em does not depend either on the
coarse-graining parameter s or on the distribution of} $\alpha$, which
in fact contribute only to $F_s$.

%This splits into two independent integrals thus obtaining
%that the integral diverges as $\langle \tau^{\gamma} \rangle$, meaning
%that $\hat{\gamma}=\mu'-1$.
%This proofs Eq. (\ref{deltamuprimo}). It is worth noticing
%that we also proved that {\em our method of observing} $\delta$
%{\em yields results that do not depend either on the coarse-graining parameter s or on the distribution of} $\alpha$.

%The slope of this trends $\alpha$ (the velocity of the rate increase)
%can be quite variable
%depending on a variety of control mechanisms.
%If the lengths of these trends make the RR time distances vary
%through several coarse graining boxes of dimension $s$ we
%write for the ``acceleration'' $\alpha$
We now establish an interesting connection with the
two-walkers model of Ref. \cite{PRE} as follows. We assume
that the trajectory $T_i$ is a zig-zag path. Each straight
line of this path is determined by the slope $\alpha_j$, which
connects the persistency time $\tau_j$ to the coarse-grain parameter
$s$, by
\begin{equation}
\alpha_j=\frac{s}{\tau_j}.
\label{alpha}
\end{equation}
If the straight line portion of the zig-zag curve has
length $l_j$, we have about $l_j \alpha_j /s$ identical values of $\tau_j$.
These are the pseudoevents of the two-walkers model of Ref. \cite{PRE}.
%In other words, the slopes (accelerations) $\alpha$ fluctuate
%slowly in time and create a time correlation among the
%identical values of $\tau_j$ determined by Eq. (\ref{alpha}).
%This model is nothing but a different way to describe
%the two walkers model of ref. \cite{PRE} in a more
%intuitive fashion. The measured probability distribution
%of $\tau_j$ reflects the dynamics of the faster walker
%while the same property for the underlying trends is connected
%to the slowest one.

\subsection{Short-term analysis}

Of course this crude model can only describe the long-time regulation
of the heartbeat dynamics and has to be supplemented with middle-
and short-time corrections so as to take into account the abrupt fall
corresponding to the first term in the r.h.s. of Eq. (\ref{cmm}).
Indeed short-term control mechanisms are due to the activity of the
autonomic nervous system which presents a very short time scale.

\begin{figure}[!h]
\centerline{\psfig{figure=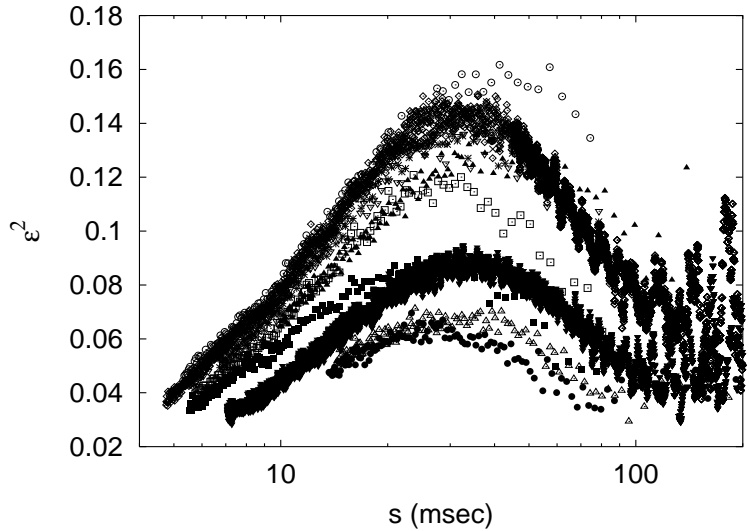,width=10.5cm}}
\caption{Short-time correlation parameter $\epsilon^2$ as a function of
the coarse-graining resolution $s$ for all the the healthy subjects. Two of them are
plotted with a much higher number of points, to show both the subtle
erratic dependence and the ``continuous'' one.}
\label{fig3}
\end{figure}

\begin{figure}[!h]
\centerline{\psfig{figure=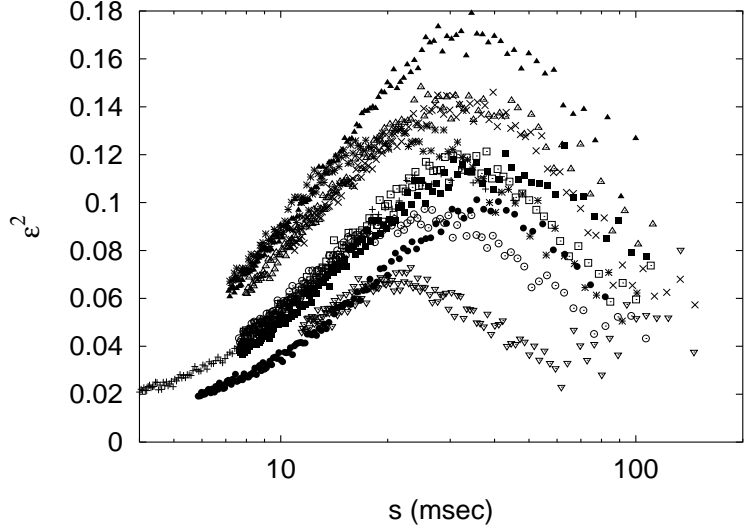,width=10.5cm}}
\caption{Short-time correlation parameter $\epsilon^2$ as a function of
the coarse-graining resolution $s$ for 5 hypertensive patients and
5 post-MI patients. We plot only half of these
groups for plot clarity.}
\label{fig4}
\end{figure}

\begin{figure}[!h]
\centerline{\psfig{figure=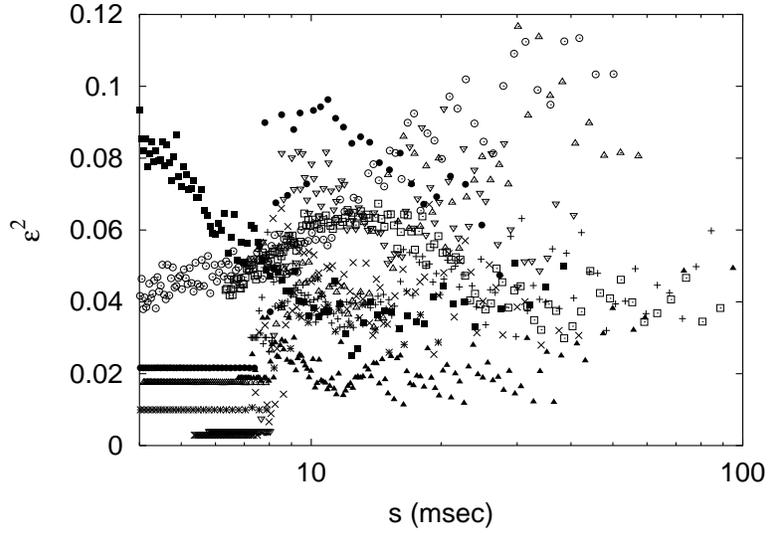,width=10.5cm}}
\caption{Short-time correlation parameter $\epsilon^2$ as a function of
the coarse-graining resolution $s$ for all 10 transplanted patients.
The structure of the previous figures is absent.}
\label{fig5}
\end{figure}

We use a short-time (local) perturbation model which does not destroy
long-time properties, in a way that is similar to the point-like 
mutations affecting the long-range correlations in DNA sequences
\cite{maria}.
The value of $\epsilon^2$ does in fact depend on the
value of $s$ and therefore we compute $\epsilon^2(s)$
in order to investigate the properties of these control mechanisms.
Surprisingly enough, the function $\epsilon^2(s)$ show a quite
universal feature for all healthy and hypertensive individuals:
we observe a steep increase as $s$ goes from few msec up to $s_{max} \cong 30$ msec
where we have a maximum. After that we have a slow decrease until $200$ msec.
The most striking result is that, even if the maximum value
is quite different among the various individuals, the values of $s_{max}$
seems to be universal, the variability among different patients being lower than $5$ msec.
This result is shown in Fig.\ref{fig3}.
This feature is somewhat visible (always for $s_{max}=30$ msec) also
in post-MI or hypertensive patients (see Fig. \ref{fig4})
and even in some CHF ones,
while it is completely absent in some CHF patients and in all the heart transplanted.
Results for these latter groups are shown in Fig. \ref{fig5}.

To the best of our knowledge,
the interpretation of $s_{max}$ is not easily explained
in physiological terms.
%On the one hand this time scale is
%so fast as to suggest a nervous interpretation;
With the due caution, we may intepret this scale as an
intrinsic resolution for the assessment of the heartbeat rate
from the control system. In other words there should be
a natural coarse graining in the physiological
determination of the interbeat time distance.
The fact that this behavior is missing in the case of 
heart transplanted patients 
seems to imply the action of the autonomic system. 
Anyway, the measured timescales rule out the 
possibility that this maximum is
due to the sampling rate of
the measuring apparatus. We conclude that more work
has to be done to explain this effect throughfully.
Finally, from a methodological point of view we use this
property to evaluate in an objective way a value
for the parameter $\epsilon^2$, which is a measure
of the {\em regularity} of the interbeat time accelerations $\alpha$ during
the trends of variation.

\begin{figure}[!h]
\centerline{\psfig{figure=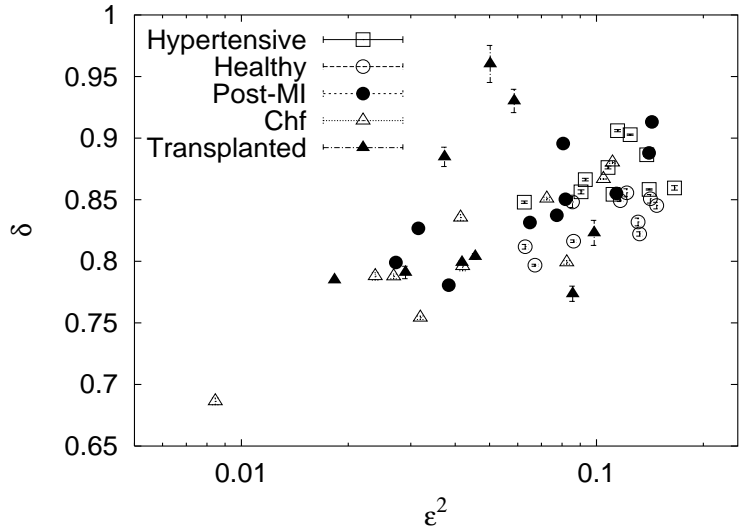,width=10.5cm}}
\caption{Bidimensional plot of $\delta(\epsilon^{2})$ for
all the 50 subjects.}
\label{fig6}
\end{figure}

At this point, as done in Ref. \cite{PRE}, we can plot
the value of $\delta$, independent of our coarse graining, as
a function of our objectively defined value of $\epsilon^2$.
The result is shown in Fig. \ref{fig6}. We recall that
both parameters reflect {\em memory}, the former being
associated with long-range fractal properties, and
the latter with short-time persistency in the heart acceleration
patterns. This latter behavior is certainly connected
with multifractal properties, as reported e.g. in Ref.
\cite{stanleyhavemercy}. Arguably, the proposed treatment
here proposed is more intuitive.

Fig. 6 shows the $\{\epsilon^2, \delta\}$ scatter plot for all 50 subjects. 
Even though there is no
satisfactory statistical discrimination among groups, maybe due to 
a variety of
factors like the limited number of patients,
differences in illness seriousness
and differences in therapy, we observe that healthy individuals 
are concentrated in a
small portion of the graph, in agreement with the results of Ref. \cite{PRE}.
Hypertensive patients, too, have their peculiar region, characterized
by an higher value of $\delta$. Other pathologies seem
to occupy broader regions. A significant portion of pathological 
individuals occupy the ``healthy zone'',
and these may
be associated to less serious conditions. Actually, preliminary 
results (unpublished) on an
enlarged group of CHF patients seem to suggest a 
correlation between serious condition and the absence/distorsion
of the $\epsilon^2 (s)$ peak at 30 msec.
These patients are also characterized
by a smaller value of $\epsilon^2 (s)$ and they are therefore
mostly outside the ``healthy'' zone.

%The main result stemming from Fig. \ref{fig6}
%is that healthy individuals are concentrated in a small
%portion of the graph, confirming the results of Ref. \cite{PRE}.
%Hypertensive patients, too, have their peculiar region, characterized
%by an higher value of $\delta$. Other pathologies seem
%to occupy broader regions. Note that a significant portion
%of pathological individuals occupy the ``healthy zone''.
%We argue that this can be associated to a less serious
%condition. Preliminary results on chf patients suggest
%a correlation between serious condition and the absence/distorsion
%of the $\epsilon^2 (s)$ peak at 30 msec. These patients are also characterized
%by a smaller value of $\epsilon^2 (s)$ and they are therefore
%mostly outside the ``healthy'' zone.

\section{Conclusions}
In conclusion, this study allows us to propose,
on the basis of a simple statistical analysis,
a series of conjectures to be accurately tested by
further studies. 
The physiological long-range memory, probably due
to a cascade of mechanisms at different timescales, can
be characterized by two parameters, the scaling $\delta$
of the underlying infinite-memory process, and the
weight $\epsilon^2$, which gives information of the
relative importance of this latter process versus
short-time mechanisms. 
The presence of
the 30 milliseconds peak of Fig. \ref{fig3}, which
allows an objective definition of $\epsilon^2$ seems to be physiological
and corresponds to the presence of a fast control
mechanism, which has to react in a time shorter than
few heartbeats to properly explain the abrupt
fall of the correlation functions. This control mechanism seems to act with
a special time resolution: a ``grid'' of about 30 milliseconds.
The absence, distorsion, or shift of this internal resolution
may reflect a pathological condition and is often associated with
a lower degree of long-time memory.

Finally, we stress that the coarse-graining procedure
adopted herein yields results that are independent of 
the coarse-graining parameter, making this analysis
robust. We believe that this analysis can be applied with
success to a variety of time series with
both long- and short-time structures, and even 
to nonstationary signals. Results stemming form our procedure 
may thereby provide additional
information useful in many diagnostic and prognostic problems.
In general, we envisage our procedure as
an automatic classifier for those complex sequences that
are not easily dealt with, using usual statistical learning methods. 
Applications may range from 
natural-language semantic classifiers
based on {\em discourse dynamics} to predictors of catastrophic events
\cite{granada}. 

\noindent {\em Aknowledgements.} 
We gratefully acknowledge financial support
from the Army Research Office through Grant DAAD 19-02-0037. One of
the authors (P. A.) acknowledges European Commission POESIA project
(IAP2117/27572) for financial support. L. P. acknowledges
ENEL Research Grant 3000021047 for support.

\end{document}